# Superradiance of entangled photon pairs from a high-density chip-scale Cs vapor cell

Heewoo Kim[1], Bojeong Seo[2], and Han Seb Moon[1,2,3]*

[1]Department of Physics, Pusan National University; Busan, 46241, Republic of Korea

[2]Quantum Science Technology Center, Pusan National University, Busan 46241, Republic of Korea

[3]Quantum Sensors Research Center, Pusan National University; Busan, 46241, Republic of Korea

Corresponding authors. Email: *hsmoon@pusan.ac.kr

**Abstract:**
Superradiance is one of the most fundamental collective quantum phenomena in light-matter interactions and has been studied extensively since Dicke's seminal work. However, its practical implementation remains challenging because superradiant enhancement requires strict experiment conditions for strong collective coupling among emitters. Can superradiance emerge in a simple platform, such as an atomic vapor cell composed of thermally moving atoms? To address this question, we identify the key signatures of superradiance in a hot atomic ensemble and find a use case of superradiance using an atomic vapor cell. Here, the Photon-Pair SuperRadiance (PPSR) process in an atomic vapor cell provides a novel approach to generating superradiant quantum light from a practical atomic platform. We experimentally demonstrate a superradiant entangled photon-pair generation via PPSR process in a high-density, 1-mm-long chip-scale Cs vapor cell. The hot, dense atomic vapor cell allows the mean interatomic distance in the Doppler-broadened atomic ensemble to be reduced to 0.29 times the idler-photon wavelength, satisfying the condition for cooperative emission. The thin chip-scale geometry enables high atomic densities while mitigating the reabsorption of emitted photons and maintaining moderate optical depth. In this subwavelength regime, we clearly observe the temporal narrowing of the biphoton wavefunction from 0.60 ns to 0.17 ns due to a superradiant decay. This pronounced temporal compression provides strong evidence of collective superradiant emission in the chip-scale Cs vapor cell. Our PPSR source delivers a detected photon-pair rate exceeding $10^6$ pairs/s while maintaining a high coincidence-to-accidental ratio (CAR) of 280. This corresponds to a nearly order-of-magnitude improvement in CAR over reported narrowband photon-pair sources at a comparable detection rate. We hope that these results establish a practical and scalable platform for realizing bright quantum light sources based on cooperative quantum effects.

## 1 Introduction

Since Dicke introduced the concept of superradiance in 1954[1], it has remained a focal point of research into cooperative emission phenomena for decades. This phenomenon has been observed in a variety of emitters, such as atoms and solid-state systems including quantum dots and nitrogen-vacancy centers in diamond[2-6]. In the single-excitation regime, superradiant light emerges directionally from an atomic ensemble coupled to a common vacuum mode, facilitated by phase-matched interactions between indistinguishable atoms[7,8]. However, practical applications remain limited because the strong collective coupling condition requires the emitters to be either confined within a distance shorter than the wavelength of the emitted light using laser cooling techniques[4], arranged at specific intervals using optical lattices[6], or controlled through additional structures such as nanohole-array apertures[4,6,9]. Consequently, these requirements have largely confined superradiance studies to sophisticated experimental systems, and realizing it on a simpler and more accessible platform, as well as finding ways to harness it for practical use, remains an open challenge.

In this regard, atomic vapor cells offer a promising alternative. They are highly accessible and versatile platforms for studying photonic quantum information technologies[10-18]. In particular, unlike cold atomic systems, the atomic vapor cells enable compact and practical implementations for quantum light sources[19], quantum memory[20], and quantum sensors[21]. In thermal atomic vapors, the atomic number density is determined by the vapor pressure, and the corresponding interatomic distance is therefore entirely set by this density. Importantly, by adjusting the cell temperature, the density of the atomic ensemble can be tuned over nearly three orders of magnitude, allowing the system to be driven continuously from a dilute, non-cooperative regime where the mean interatomic distance exceeds the optical wavelength to a subwavelength cooperative regime where collective emission effects become prominent. Despite this advantage, superradiant decay has remained difficult to observe in hot atomic vapors. Because increasing the atomic density also increases the reabsorption of emitted photons and limits photon extraction from the medium[22,23], these effects become more severe as the system approaches the cooperative regime. Conversely, at low density the photon-pair rate is too low to characterize the emission reliably. As a result, clear superradiant decay has not been reported in a dense vapor cell.

In this work, we demonstrate superradiance in a 1-mm-long Cs vapor cell fabricated using MEMS technology. The thin geometry allows a high atomic density while reducing photon reabsorption and maintaining moderate optical depth (OD). As a result, the mean interatomic distance can be reduced below the emitted-photon wavelength, satisfying the condition for cooperative emission. In the cascade-type spontaneous four-wave mixing (SFWM) process, a signal photon is generated first and is followed by the emission of an idler photon. The signal photon prepares a collective atomic excitation that drives the subsequent decay of the idler mode. This process can be viewed as a triggered form of superradiance. We refer to this collective two-photon emission process as photon-pair superradiance (PPSR). We observe a pronounced narrowing of the biphoton temporal waveform as the mean interatomic distance decreases below the photon wavelength. This behavior is consistent with collective enhancement of the two-photon decay process. The superradiant emission also increases the photon-pair generation rate while preserving strong quantum correlations. These results extend superradiance to entangled photon-pair generation in a hot atomic ensemble. Our approach provides a practical platform for exploring superradiant phenomena and their applications.

# 2 Results

## 2.1 Photon-pair superradiance process

Many previous superradiance phenomena are commonly described for dipole-allowed transitions of two-level atoms[2-6]. However, superradiant emission is not limited to simple two-level transitions. In a three-level cascade-type two-photon transition, the collective effect of photon-pair generation can emerge due to two-photon atomic coherence[22,23]. Collective two-photon coherence induced by two-photon-resonant excitation can make a cascade-type inhomogeneously broadened atomic ensemble behave as an effectively homogeneous medium[24]. In a highly dense atomic medium, where the mean interatomic distance is subwavelength, the entangled photon pairs undergo a collective decay.

Figure 1 shows the PPSR process in our cascade-type SFWM scheme, which we describe step by step. The first step, involving the emission of the signal photon, is initiated by an adiabatic excitation that prepares the ensemble in a Dicke-like state. The single atomic excitation in a collection of atoms in the $6P_{3/2}$ intermediate state is conditionally prepared as

$$|\Psi\rangle = \sum_{j}^{N} C_j e^{i\delta_j t} e^{i\vec{k}_I \cdot \vec{r}_j} \left| g_1, \cdots, i_j, \cdots, g_N \right\rangle |0\rangle, \tag{1}$$

where $N$ is the number of atoms, $\vec{k}_I$ is the wave-vector of the idler field, $\delta_j$ is the detuning of the idler field, $\vec{r}_j$ is the position of j-th atom in the ensemble, and $C_j$ is the probability amplitude of finding atoms in the intermediate state. In this $N$-atom entangled state, a single atomic excitation is shared among a large number of atoms.

In the second step, the idler photon is spontaneously emitted via a superradiant decay. The emission of signal photon, adiabatically seeded by a pre-established two-photon coherence, plays the role of a trigger photon that initiates the superradiant decay ($\Gamma_{SR}$) of the idler photon via the PPSR process. Consequently, the probability amplitudes of detecting an idler photon in the phase-matched direction are coherently superposed over the spatial and velocity distributions of atoms, leading to directional enhancement via constructive interference.

The experimental configuration for entangled photon pair generation via SFWM in cascade-type atomic system is shown in the inset of Fig. 1a. The pump and coupling lasers interact with the $6S_{1/2}$–$6P_{3/2}$ and $6P_{3/2}$–$6D_{5/2}$ transitions of $^{133}$Cs, respectively, with detunings $\delta_P$ and $\delta_C$ from the corresponding resonances, and a two-photon detuning $\delta_{two} = \delta_P + \delta_C$. The signal photon is generated and precedes the emission of idler photon. As is well known, two-photon resonance occurs via the application of counter-propagating pump and coupling lasers satisfying the Doppler-free two-photon resonant condition in the hot atomic ensemble. Spatially well-overlapping pump and coupling lasers interact with a 1-mm-long, thin Cs vapor cell, generating PPSR from a high-density atomic vapor whose density can be tuned by adjusting the cell temperature.

Figure 1b quantitatively shows how the mean interatomic distance and OD evolve with temperature for vapor cells of different lengths. Increasing the cell temperature raises the atomic number density, thereby reducing the interatomic distance. For sufficiently short cells, the system enters the subwavelength regime while maintaining moderate OD, providing favorable conditions for observing superradiant emission of entangled photon pairs. Here, the subwavelength regime ( $r_{SR} < \lambda_I$ ) denotes the onset of cooperative effects, whereas the superradiant regime corresponds to the pronounced collective decay observed for $r_{SR} < \lambda_I/2$.

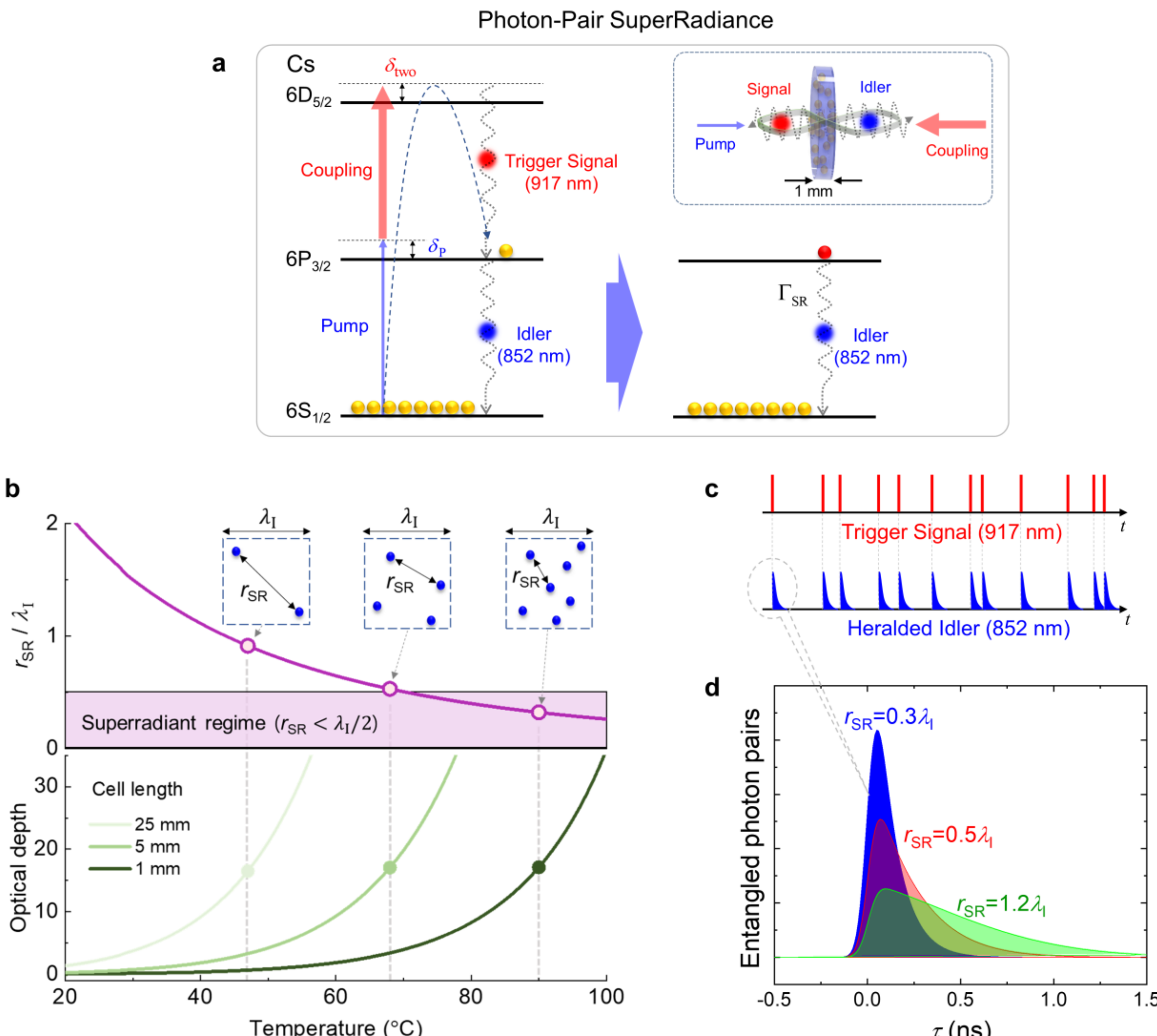

**Fig. 1** Experimental configuration for photon-pair superradiance process. **a** Photon-pair superradiance process in three-level cascade-type atomic system. The signal photon plays the role of a trigger photon and the adiabatic excitation prepares the ensemble in a Dicke-like state. The idler photon is then spontaneously emitted through the superradiant decay. The inset illustrates the experimental scheme. **b** Temperature dependence of the mean interatomic distance and optical depth for Cs vapor cells of different lengths. For shorter cells, the system reaches the superradiant regime while maintaining moderate optical depth, enabling cooperative emission. **c** The signal photons are generated under continuous excitation and exhibit thermal photon statistics. When the triggering signal photon defines the start time (gray dashed lines), the idler photons appear in a pulsed temporal mode with a superradiance. **d** The biphoton wavefunction exhibits pronounced temporal narrowing as the mean interatomic distance decreases ( $r_{SR} = 0.3\lambda_I$, $0.5\lambda_I$, $1.2\lambda_I$ ). Superradiance of the entangled photon pair is observed when $r_{SR} < \lambda_I/2$.

The PPSR can be verified by measuring the second-order cross-correlation between the signal and idler photons. Fig. 1c schematically illustrates the signal and idler photons emitted and detected under continuous-wave pump and coupling lasers. Using the signal photon as a heralding photon allows us to measure how rapidly the subsequent idler photon is emitted. Fig. 1d shows the coincidence histogram as a function of the detection-time difference between the two photons. When superradiance occurs, the histogram is expected to narrow as the mean interatomic distance becomes smaller than the idler wavelength.

To describe this temporal narrowing quantitatively, we consider the biphoton temporal wavefunction of the PPSR process in the hot atomic ensemble. For an atom with velocity $v$, the wavefunction can be written as $\Psi_v(\tau) \equiv \langle 0 | \hat{a}_\mathrm{I}(t_\mathrm{I}) \hat{a}_\mathrm{S}(t_\mathrm{S}) | \Psi_v \rangle$, where $\hat{a}_\mathrm{S}(t_\mathrm{S})$ and $\hat{a}_\mathrm{I}(t_\mathrm{I})$ denote the annihilation operators for the signal and idler photons measured at times $t_\mathrm{S}$ and $t_\mathrm{I}$, respectively, and $\tau = t_\mathrm{I} - t_\mathrm{S}$ is the detection-time difference. Considering the superradiant decay rate of the idler-photon mode, $\Psi_v(\tau)$ can be described as follows

$$\Psi_v(\tau) = A(v) e^{(-\Gamma_\mathrm{SR}/2 + i k_\mathrm{I} v)\tau}, \tag{2}$$

where $k_\mathrm{I}$ is the idler-photon wavevector, and $A(v)$ is a velocity-dependent complex amplitude that quantifies the degree of two-photon coherence generated in each velocity group (see Sec. II in the Supplemental Material). The superradiant decay rate $\Gamma_\mathrm{SR}$ of the idler mode is described by

$$\Gamma_\mathrm{SR} = \Gamma_\mathrm{I}(1 + \mu N), \tag{3}$$

where, $\Gamma_\mathrm{I}$ is the intrinsic decay rate of the idler photon mode of the $6S_{1/2}$–$6P_{3/2}$ transition and $\mu$ denotes the geometric factor[25], assuming emitters distributed within a cylindrical volume (see Sec. III in the Supplemental Material).

The superradiant enhancement depends critically on the mean interatomic distance $r_\mathrm{SR}$, which is governed by the temperature of the atomic vapor cell. The number of atoms $N$ within the interaction volume, and the corresponding $r_\mathrm{SR}$ are calculated from the temperature-dependent vapor pressure of cesium atoms[26]. Using the known relation between vapor pressure and atomic number density, $N$ and $r_\mathrm{SR}$ can be written as

$$r_\mathrm{SR} = \frac{5}{9} \left(\frac{N}{V}\right)^{-\frac{1}{3}}, \tag{4}$$

where $V$ is the interaction volume. As the temperature of the vapor cell increases, the atomic density increases, so $r_\mathrm{SR}$ tends to decrease. A detailed table summarizing the calculated values of cell temperature, optical depth, and $r_\mathrm{SR}$ is provided in Supplemental Material Sec. I. Substituting Eq. (4) into Eq. (3) yields the superradiant decay rate as follows:

$$\Gamma_\mathrm{SR} = \Gamma_\mathrm{I}\left[1 + \mu \frac{V}{\lambda_\mathrm{I}^3} \left(\frac{9}{5} \frac{r_\mathrm{SR}}{\lambda_\mathrm{I}}\right)^{-3}\right]. \tag{5}$$

The second-order cross-correlation function $g_{\mathrm{SI}}^{(2)}(\tau)$ for the entangled photon pairs generated from a Doppler-broadened atomic ensemble is defined as[27,28]

$$g_{\mathrm{SI}}^{(2)}(\tau)=\left|\int \Psi_{v}(\tau) f(v) d v\right|^{2} H(\tau), \tag{6}$$

where $f(v)$ is a one-dimensional Maxwell-Boltzmann velocity distribution function, and $H(\tau)$ is the Heaviside step function. The temporal shape of Eq. (6) corresponds to the temporal biphoton wavefunction, related to the temporal behavior of the PPSR process. Here, because $\Psi_{v}(\tau)$ of Eq. (2) includes the superradiant decay rate of the idler photon mode, the temporal width of $g_{\mathrm{SI}}^{(2)}(\tau)$ becomes narrower as $\Gamma_{\mathrm{SR}}$ increases. Overall, a higher cell temperature reduces $r_{\mathrm{SR}}$ and enhances the superradiance, which narrows the temporal width of $g_{\mathrm{SI}}^{(2)}(\tau)$. This temporal narrowing both confirms the PPSR process and provides a quantitative measure of its strength.

### 2.2 Superradiance of entangled photon pairs

We experimentally investigate how the temporal waveform of the entangled photon pairs depends on $r_{\mathrm{SR}}$. As the cell temperature increases from 21°C to 95°C, $r_{\mathrm{SR}}$ decreases from $2\lambda_{\mathrm{I}}$ to $0.29\lambda_{\mathrm{I}}$. To quantify the superradiance strength, we define the normalized temporal probability distribution of the entangled photon pairs as

$$P_{\mathrm{I}}(\tau)=\frac{g_{\mathrm{SI}}^{(2)}(\tau)}{\int d\tau\, g_{\mathrm{SI}}^{(2)}(\tau)}, \tag{7}$$

which gives the temporal probability of detecting heralded idler photon.

Figure 2a shows the measured $P_{\mathrm{I}}(\tau)$ at different vapor cell temperatures. At 21 and 37 °C, where $r_{\mathrm{SR}} \geq \lambda_{\mathrm{I}}$, the two temporal waveforms are nearly indistinguishable. As the cell enters the subwavelength regime, $r_{\mathrm{SR}}<\lambda_{\mathrm{I}}$, the temporal peak becomes significantly narrower, indicating the onset of superradiance.

We further analyse the temporal width of $P_{\mathrm{I}}(\tau)$ as a function of the ratio of $r_{\mathrm{SR}} / \lambda_{\mathrm{I}}$, as summarized in Fig. 2b. The inset in Fig. 2b shows the superradiance strength $\Gamma_{\mathrm{SR}} / \Gamma_{\mathrm{I}}$, extracted by comparing the measured temporal widths with the theoretical linewidth obtained from Eq. (6) after convolution with 100 ps detection timing jitter (see Sec. IV in the Supplemental Material). As the cell temperature increases and $r_{\mathrm{SR}}$ decreases, the temporal width of $P_{\mathrm{I}}(\tau)$ narrows significantly from 0.60 ns to 0.17 ns. In the regime of $r_{\mathrm{SR}}<\lambda_{\mathrm{I}}$, a clear compression of the temporal width of $P_{\mathrm{I}}(\tau)$ is observed, which becomes more pronounced as $r_{\mathrm{SR}}$ approaches and drops below $\lambda_{\mathrm{I}} / 2$.

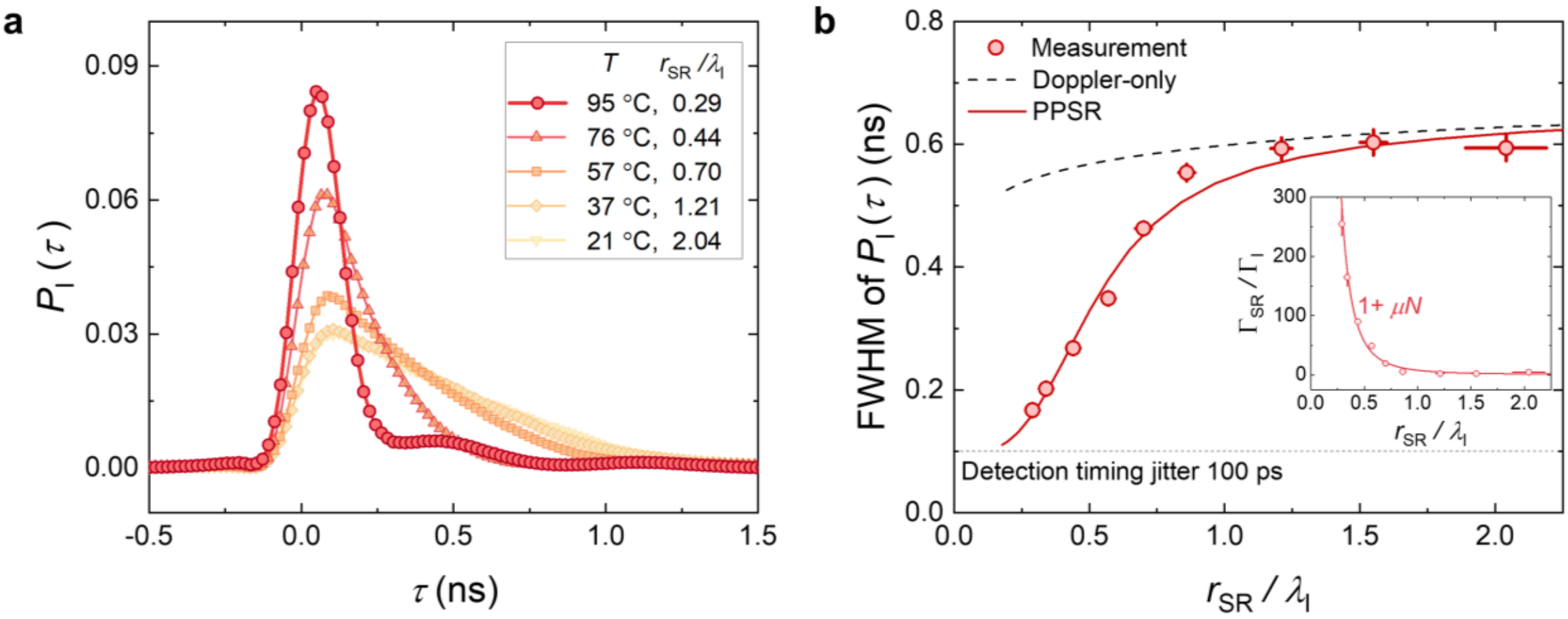

**Fig. 2** Superradiance of entangled photon pairs. **a** Normalized temporal probability distribution $P_I(\tau)$ of the heralded idler photon measured at different vapor cell temperatures. Increasing temperature leads to clear superradiant temporal narrowing. **b** Temporal width (FWHM) of the entangled photon pairs as a function of the ratio $r_{SR}/\lambda_I$. The experimental data (red circles) are in good agreement with the PPSR model (red solid curve), whereas the Doppler-only model (black dashed curve) fails to reproduce the observed temporal narrowing. The gray dotted line denotes the 100 ps detection timing jitter. The inset shows the extracted superradiance strength.

The measured FWHM of $P_I(\tau)$ are compared with two theoretical models. The first is the Doppler-only model (black dashed curve in Fig. 2b), which neglects collective enhancement ($\Gamma_{SR} = \Gamma_I$ in Eq. (3)). In this model, the width of $P_I(\tau)$ decreases with the Doppler velocity distribution, but at high temperature it differs markedly from the measured values. The second is the PPSR model (red solid curve), calculated from Eq. (7) with the superradiant decay rate of Eq. (3) and the detection timing jitter. This model, which incorporates the temperature-dependent increase of the decay rate, agrees well with the experimental data. According to Eq. (7), the superradiant idler emission overcomes the width predicted by Doppler-only model when the decay rate exceeds about $50\Gamma_I$. Such a large enhancement of the decay rate would be difficult to attribute to other temperature-dependent mechanisms, such as collisional broadening[29]. Above all, since the temperature dependence of the narrowing is well reproduced by the PPSR model, the reduction in temporal width provides clear evidence of the PPSR process in our cascade-type four-wave-mixing system.

### 2.3 Superradiant enhancement via entangled photon-pair superradiance

Owing to superradiance in the cascade-type PPSR process, our dense 1-mm-long Cs vapor cell yields a highly bright photon-pair source. Fig. 3a shows the entangled photon pairs counts as a function of OD by controlling the cell temperature, at a pump power of 0.03 mW and a coupling power of 15 mW. The experiment results are directly measured over 30 s and each histogram bin corresponds to 20 ps. For clarity, the OD = 20 curve is shown at its actual scale, while the remaining curves are magnified by ×5, ×30, ×500, and ×10000 in order of decreasing OD. As the temperature increases, the number of participating atoms per unit volume and the superradiance rise and the brightness of the PPSR source is enhanced. At the same time, we can see that stronger superradiant coupling induces faster collective emission dynamics, resulting in an increased CAR.

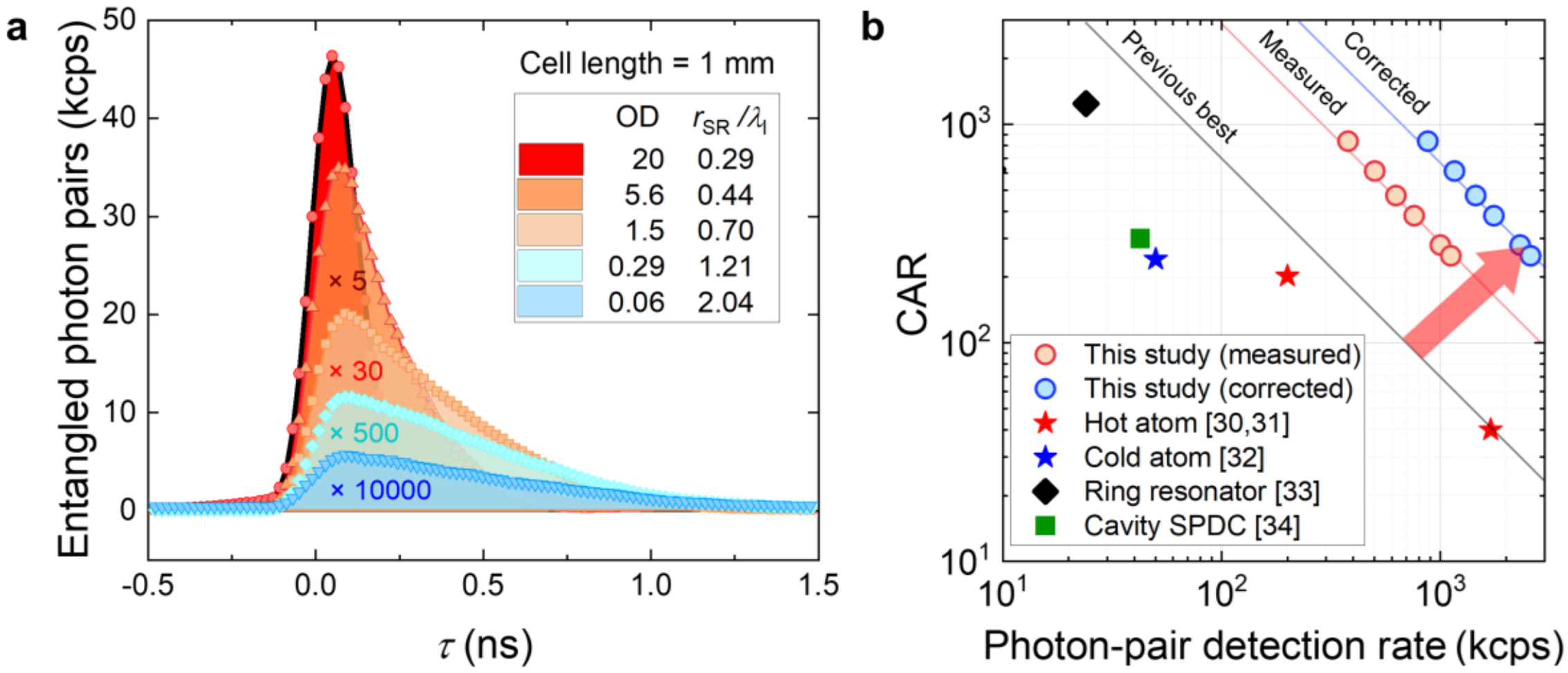

**Fig. 3** Brightness enhancement via entangled photon-pair superradiance. **a** Entangled photon pairs count rate measured at different Cs vapor cell temperatures. Each curve corresponds to a distinct temperature, with the associated OD and mean interatomic distance $r_{SR} / \lambda_I$ indicated in the legend. For clarity of presentation, the lower OD curves are multiplied by ×5, ×30, ×500, and ×10000. **b** Photon-pair detection rate versus CAR, including results from different narrowband photon-pair generation platforms. The red circles represent measured results and the blue circles indicate the expected performance assuming improved detector quantum efficiency.

A higher OD enhances the photon-pair source. Indeed, after accounting for the scaling factors in Fig. 3a, the photon-pair counts increase markedly from OD = 5.6 to OD = 20. At the same time, a higher OD increases the reabsorption loss of the idler photons. These two competing effects give rise to an optimal OD[30], which increases as the cell length decreases, reaching approximately 4 for 25 mm[30], 12 for 5 mm[31], and 20 for a 1 mm cell used here (see Sec. V in the Supplemental Material). The thin vapor cell is particularly advantageous in this regard. As shown in Fig. 1b, for a given OD a shorter cell reaches a smaller mean interatomic distance, and thus a stronger superradiant regime, than a longer one. This allows the thin cell to access strong superradiance while keeping the OD moderate, so that the enhanced collective decay rate (Fig. 2b, inset) broadens the idler spectrum beyond the atomic absorption linewidth and a smaller fraction of idler photons is lost to reabsorption. The suppressed reabsorption loss shifts the optimum toward higher OD, enabling bright photon-pair generation.

Figure 3b compares our photon-pair source with previously reported narrowband (spectral linewidths below a few gigahertz) photon-pair systems. The plot shows the photon-pair detection rate versus CAR to provide a practical comparison based on directly measurable quantities that inherently include optical losses. The red circles represent the raw data obtained with a 15 mW coupling laser while varying the pump power from 0.03 to 0.09 mW, where the quantum efficiencies of the single-photon detectors are 50% and 70% for the signal and idler photons, respectively. For a pump power of 0.08 mW, the corresponding CAR of 280 yields a photon-pair detection rate of $10^6$ pairs/s. Other symbols correspond to reported results from four-wave mixing in hot atomic vapors[30,31], cold-atom ensembles[32], micro-ring resonators[33], SPDC in optical cavities[34], respectively.

For a fair comparison, we also show our results corrected to a detection efficiency of 90% for both the signal and idler photons (blue circles). The solid lines (gray, red, and blue) in Fig. 3b trace the relationships between CAR and photon-pair detection rate, with the gray line from the previously reported best-performing results and red and blue lines from our measured and corrected data. From this comparison, our PPSR-based source establishes a new state-of-the-art performance for narrowband entangled photon-pair source. To further characterize the entanglement nature of the photon pairs, we performed quantum state tomography on the polarization-entangled state generated by the PPSR source. The measured fidelity to the maximally entangled Bell state reaches 0.95, confirming that superradiant emission preserves high-quality polarization entanglement.

## 3 Discussions

We have demonstrated entangled photon-pair superradiance in a dense, thin chip-scale $^{133}$Cs vapor cell via spontaneous four-wave mixing. Despite the Doppler-broadened nature of the thermal atomic ensemble, the mean interatomic distance was reduced into the subwavelength regime, enabling the observation of superradiant decay and the resulting temporal narrowing of the biphoton wavefunction to 0.17 ns. In the subwavelength regime, after detection efficiency correction, the photon-pair rate achieves $2.3\times10^6$ pairs/s with CAR of 280. These results demonstrate that our PPSR-based source yields a CAR nearly an order of magnitude higher than the reported measured values of previous narrowband implementations at a comparable photon-pair detection rate, as shown in Fig. 3b, highlighting the practical potential of superradiance for quantum photonic applications.

More broadly, our work extends the concept of superradiance to entangled photon-pair generation in a hot atomic ensemble. The demonstration of PPSR in a simple chip-scale atomic vapor cell provides a practical use case for superradiance in nonlinear quantum optics. Operating in continuous-wave mode at room temperature, this system allows us to easily control the mean interatomic distance just by adjusting the cell temperature. The high brightness, in turn, provides a high sampling rate that enables fast superradiance experiments. Moreover, the modified collective decay rate induced by PPSR alters the spectral bandwidth of the photon, and this tunability could be exploited to spectrally match photon-pair sources to emitters on other platforms[35]. We anticipate that these findings will stimulate further research at the interface of atomic physics, quantum optics, and quantum information science.

## 4 Methods

## Experimental setup

The experiment for PPSR was performed using the 1-mm-long Cs vapor cell. The vapor cell outer dimensions are 2 mm × 3.5 mm × 1.4 mm. The diameter and thickness of the photon-pair chamber were 1.5 mm and 1 mm, respectively. The pump and coupling beams were counter-propagating and overlapped in the center of the cell. The pump and coupling lasers are orthogonally linearly polarized along the horizontal and vertical polarization directions, respectively.

Owing to the conservation of angular momentum in the SFWM process, the generated photon pair in the two-photon decay is polarization-correlated with the perpendicular linear-polarized signal/idler photons. A continuous-wave pump laser at 852 nm was blue-detuned from the $6S_{1/2}(F=4) \rightarrow 6P_{3/2}(F'=5)$ transition by $\delta_P = 1.31$ GHz. The coupling laser at 917 nm was red-detuned from the $6P_{3/2}(F'=5) \rightarrow 6D_{5/2}(F''=6)$ transition by $\delta_C = -1.35$ GHz. The resulting two-photon detuning was $\delta_{two} = -40$ MHz. The pump power was varied between 0.03 mW and 0.09 mW, whereas the coupling power was kept constant at 15 mW.

The generated signal and idler photons pass through a band pass filter and polarizer to remove residual laser components, after which they are coupled into a single-mode fiber and delivered to the detectors. For photon detection, we employ superconducting nanowire single-photon detectors (SNSPDs) optimized for 780 nm. The quantum efficiencies of detectors are 70 % for 852 nm, and 50% for 917 nm, respectively, and the timing jitter of the detector is approximately 70 ps.

## Supplementary Information

The online version contains supplementary material available at XX.

Supplementary material 1.